# Direct strain and slope measurement using 2D DSPSI Title

Wajdi DANDACH[1, a]**,** Jérôme MOLIMARD[2,b] and Pascal PICART[3,c]

[1]LTDS, UMR CNRS 5513, Université de Lyon, Ecole des Mines de Saint-Etienne, Centre SMS, 158 cours Fauriel, 42023 Saint-Etienne cedex 02, France

[2]UMR CNRS 5146, Université de Lyon, Ecole des Mines de Saint-Etienne, Centre CIS,158 cours Fauriel, 42023 Saint-Etienne cedex 02, France

[3]LAUM, UMR CNRS 6613, ENSIM, Université du Maine,

Rue Aristote - 72085 Le Mans cedex 09, France

[a]dandach@emse.fr, [b]molimard@emse.fr, [c]pascal.picart@univ-lemans.fr

**Abstract.** Large variety of optical full-field measurement techniques are being developed and applied to solve mechanical problems. Since each technique possess its own merits, it is important to know the capabilities and limitations of such techniques. Among these optical full-field methods, interferometry techniques take an important place. They are based on illumination with coherent light (laser). In shearing interferometry the difference of the out of-plane displacement in two neighboring object points is directly measured. Since object displacement does not result in interferometry fringes, the method is suited for localization of strain concentrations and is indeed used in industry for this purpose. Used quantitatively DSPSI possesses the advantage over conventional out-of-plane displacement-sensitive interferometry that only a single difference of the unwrapped phase map is required to obtain flexural strains, thereby relieving problems with noise and reduction in the field of view. The first publication [1] on (DSPSI) was made in 1973, but the emergence of a system providing quantitative measurements is more recent [2].

This work aims to present the results of strain measurements using digital speckle pattern shearing interferometry (DSPSI).

## Introduction

We used an optical full-field technique (OFFT) for measuring surface strain fields of composite coupon submitted to bending and shear in a modified Iosipescu device. Most of OFFT measure displacements (Electronic Speckle Patter Interferometry, Grid Method, Moiré Interferometry and so on); one of the challenging parts of displacement measurement techniques is the calculation of the strain map from the displacement map. In particular, noise propagation and lens distortion has to be carefully treated. Adjacent to the methods measuring displacements, DSPSI measures directly displacement derivatives of surfaces. More accurately, it eliminates the reference beam of holographic or speckle interferometry, which leads to a simplified optical set-up, not requiring special vibration isolation. These advantages have exhibited practically DSPSI a surface strain and rotation measurement system. Historically, main drawback of the technique is the use of an image doubling element, and a loose of spatial resolution. Since this technique combines temporal phase shifting with the use of a small shear distance (50μm in this experiment) [3], it is shown here that it is possible to find a very good compromise between resolution, spatial resolution and sensitivity.

**Experimental Conditions**

**Optical Set-up.**

The set-up used for this experiment is optimised to reduce signal upon noise ratio and to maximise spatial resolution (Fig. 1). A laser ($\lambda$=532 nm) illuminates the front surface of the specimen [4]. Light is sequentially injected in the four optical fibers by using an optical switch. The outputs fibers are attached to a device, manufactured in the laboratory, which allows the laser collimated beam illuminates the surface of the specimen by four directions (Fig.1). This way, the four necessary directions of illumination can be realized, with equivalent illumination angle. The diffracted beam from the specimen is sheared by Michelson interferometer containing a beam splitter and two mirrors. First mirror is fixed; the second is controlled by a 3-axis PTZ device PSH 1z NV from Piezo-system Jena, capable of tilting or translating the mirror. As follow, we can realize the shear in x or y directions by tilting one mirror, similarly the piston movement for the temporal phase stepping, we can observe images.

Acquisition is performed using a 12 bit Jai camera equipped with a zoom lens, connected to a gigabit Ethernet card plugged into a PC. Two images can be observed due to shearing distance, these two images are recovered by a Labview program that controls shear distance and performs their difference. The same is done for the deformed state. As a result two images are obtained and named respectively reference image and deformed image. These two images are called by a Matlab program to process image data.

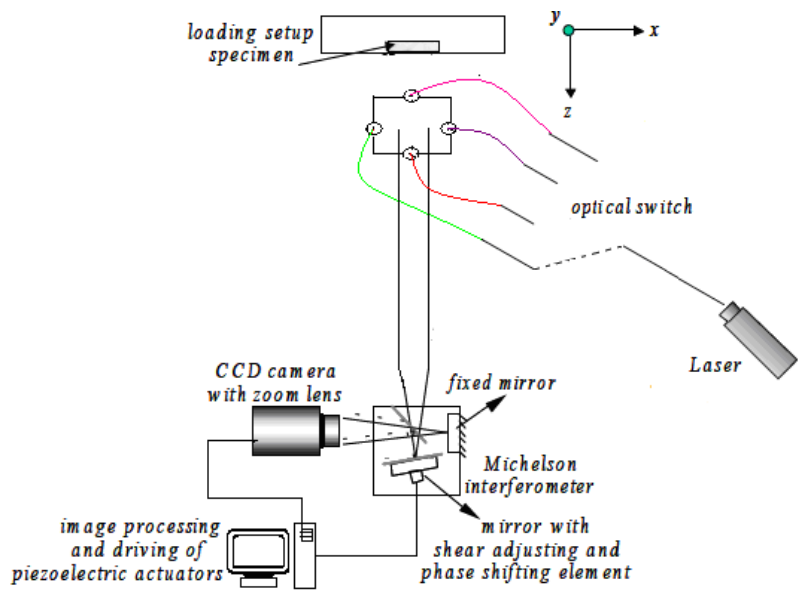

**Fig. 1 : DSPSI Set-up**

**Mechanical conditions:**

The Iosipescu mechanical set-up based on the EMSE fixture [5] is presented in Fig 2. Measurements are performed in the middle of the specimen (grid lines). On the movable jaw we imposed a vertical displacement. A classical load cell is used to control the load. The specimen has been loaded at 675 N. for this experiment; the sample width is 20 mm and its thickness 3 mm, the distance between the two clamps being 17.9 mm.

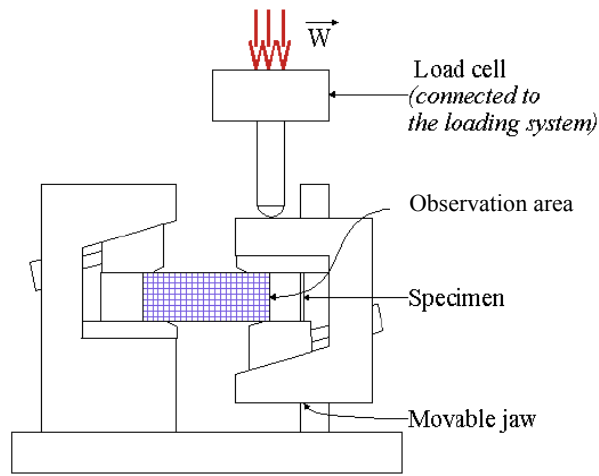

**Fig. 2 :** Iosipescu Mechanical set-up

**Experimental procedure**

During the Iosipescu test, the surface strains of specimen induced by external loads have been digitised in the form of phase maps corresponding to the changes on the attached specimen. At each load step, eight phase difference maps for the combination between 4 directional illuminations and 2 applied shear directions have been determined. Ten intensity maps obtained by a phase stepping technique enable to calculate one phase map for each of the eight combinations. Subsequently, Matlab calculated the phase change maps for each combination between two different load steps. In this calculation, no fitting or filtering methods are used, thus spatial resolution remains maximum. Then, in-plane strains and out-of plane displacement derivatives have been analysed at each load step using Matlab. Finally, salt and pepper filtering was applied to in-plain strain maps and out-of-plane displacement derivative maps to increase the map visibility. In the end, all filtered images with physical information have a spatial resolution of about 1 mm. However, the original image before filtering keeps the spatial resolution of about 0.114 mm.

Results will be presented during the conference.

**Conclusion**

This paper presents an original set-up for the direct measurement of strain map using quantitative DSPSI. Application to the Iosipescu shear test demonstrated the feasibility of the measurement. Further work will focus on strain map optimization and metrology.